\documentclass[prd,preprint,showpacs,preprintnumbers,nofootinbib,superscriptaddress]{revtex4}

\usepackage{amsmath}
\usepackage{amssymb}
\usepackage{graphicx}
\usepackage{txfonts}
\usepackage{color}

\newcommand{\sla}[1]{{/\hskip-0.55em{}#1}{}}
\newcommand{\SU}{\textrm{SU}}
\newcommand{\SO}{\textrm{SO}}

\definecolor{gray}{gray}{0.5}

\begin{document}


\preprint{RIKEN-TH-66}
\preprint{hep-th/0602244}

\title{%
Stability of 4-dimensional Space-time from IIB Matrix Model \\ via 
Improved Mean Field Approximation }


\author{T.~Aoyama}
\affiliation{Theoretical Physics Laboratory, RIKEN, Wako, 351-0198, Japan }

\author{H.~Kawai}
\affiliation{Theoretical Physics Laboratory, RIKEN, Wako, 351-0198, Japan }
\affiliation{Department of Physics, Kyoto University, Kyoto, 606-8502, Japan }

\author{Y.~Shibusa}
\affiliation{Theoretical Physics Laboratory, RIKEN, Wako, 351-0198, Japan }

\begin{abstract}
The origin of our four-dimensional space-time has been pursued 
through the dynamical aspects of the IIB matrix model via the improved 
mean field approximation. 
Former works have been focused on the specific choice of configurations 
as \textit{ansatz} which preserve $\SO(d)$ rotational symmetry. 
In this report, an extended ansatz is proposed and examined up to 
3rd order of approximation which includes both $\SO(4)$ ansatz and 
$\SO(7)$ ansatz in their respective limits. 
From the solutions of self-consistency condition represented by 
the extrema of free energy of the system, it is found that a part 
of solutions found in $\SO(4)$ or $\SO(7)$ ansatz disappear in 
the extended ansatz. 
It implies that the extension of ansatz works as a device to 
distinguish the stable solutions from the unstable ones. 
It is also found that there is a non-trivial accumulation of 
extrema including the $\SO(4)$-preserving solution, which may lead 
to the formation of \textit{plateau}. 
\end{abstract}





\maketitle

\section{Introduction and Summary}

The IIB matrix model has been proposed as a constructive formulation 
of superstring theory
\cite{Ishibashi:1996xs,Aoki:1998bq}.
One of the significant features of the model is that the 
space-time itself is expressed by the eigenvalue distributions of 
10 bosonic matrices, and thus treated as a dynamical variable. 
Therefore, the origin of our four-dimensional universe can be 
argued on the basis of this framework as a spontaneous symmetry 
breakdown of Lorentz symmetry
\cite{Ambjorn:2000dx,Hotta:1998en,Nishimura:2004ts,Nishimura:2001sx,Kawai:2002jk,Kawai:2002ub,eighth}.
The subject has been pursued up to now with a technique called 
the improved mean field approximation (IMFA)
\cite{Oda:2000im,Sugino:2001fn,Kabat:1999hp}.
By the systematic application of the method, it was revealed that among 
$\SO(d)$-preserving configurations $d=4$ case seems to 
be preferred and that the extent of space-time turns out to be large 
for the four-dimensional directions 
against the rest of six-dimensional part. 
This result suggests the spontaneous breakdown of Lorentz symmetry 
and the emergence of our four-dimensional space-time
\cite{Nishimura:2001sx,Kawai:2002jk,Kawai:2002ub,eighth}.

The IMFA method is a systematic improvement of variational method. 
It introduces quadratic terms to the original action, 
which may be considered as artificial mean fields, 
and formulate a perturvative series expansion, 
which involves 
a number of parameters as coefficients 
of those quadratic terms. 
We then reorganize the series by resummation 
to obtain the forms of improved series. 
The determination of parameters is guided by the principle of 
minimal sensitivity \cite{Stevenson:1981vj}; 
the result should be least sensitive to those nominal parameters. 
It is found in several examples that the improved series is  
stable in some regions of parameters, in which the dependences on 
those parameters 
would be considered to vanish effectively, and the {\em exact} value 
should be reproduced. 
Thus, the solution to the consistency condition is formulated as 
identifying such a flat region. 
We denote it as {\em plateau}. 

The analyses of the IIB matrix model based on the IMFA method 
have been done only for a restricted set of configurations. 
It becomes an enormous task to solve the above consistency conditions 
because there is quite a large number of parameters 
introduced along the prescription of the IMFA method. 
In order to reduce the number of parameters and to clarify the 
physical implications, restrictions were imposed on the set of 
parameters so that the $\SO(d)$ subgroup of $\SO(10)$ symmetry 
stays intact. 
In particular, $d=4$ and $7$ cases have been examined intensively. 
At this stage 
we have to reflect whether or not those choices of configurations 
are reasonable and proper. 
In the present report we examine a wider set of parameters 
which covers both $\SO(4)$ and $\SO(7)$ cases as its subsets. 
It is referred as ``4-3-3'' ansatz below. 
With this particular ansatz we can treat the solutions of both 
$\SO(4)$ and $\SO(7)$ ansatz on equal footings 
and discuss the problem which vacuum would be more preferred. 

A wider class of configurations also sheds lights on the 
problem whether or not the plateau is actually realized 
in the analysis thus far carried out up to high orders. 
This issue may be translated to the following statement. 
If the IMFA were to realize the minimal sensitivity, 
the value on plateau should be independent of any nominal parameters. 
Then, plateau must be stable even if we adopt a wider class of ansatz. 

We evaluated up to third order contributions of improved series 
and found the solutions of consistency conditions defined by 
the extrema of the improved free energy with respect to 
the artificial parameters. 
It turned out that some of the solutions of $\SO(4)$ ansatz 
and $\SO(7)$ ansatz also appear in the extended parameter space, 
while the others vanish. 
The latter set of extrema are considered to be ``unstable'' 
in the sense that they correspond to the saddle points 
of the improved series. 

There is found a non-trivial accumulation 
of extrema including one of $\SO(4)$-preserving solutions, 
which has been seen in the higher order calculations to belong 
to the (would-be) plateau in $\SO(4)$ ansatz 
\cite{Kawai:2002ub,eighth}. 
Therefore, this accumulation will give an affirmative support on 
the plausibility that the solution found in $\SO(4)$ ansatz 
forms a plateau. 
It implies that the four-dimensional universe is more preferred  
to the seven-dimensional universe. 

It is noted along with the present analysis that 
the extension of artificial parameter space is as efficient as 
the calculation of higher order perturbation in the IMFA method. 
In the former analyses of $\SO(4)$ and $\SO(7)$ ansatz, 
higher-order calculations were required to obtain fair signal 
for development of plateau.

\section{Improved Mean Field Approximation}

The model we examine in this report is the IIB matrix model defined by 
the partition function with the action $S$ as 
\begin{equation}
	Z = \int\!dA\,d\psi\ e^{-S} \,,
\end{equation}
\begin{equation}
	S = N\,{\rm Tr}\,\Biggl[
		-\frac{\lambda}{4} [ A_\mu, A_\nu ]^2 
		- \frac{\sqrt\lambda}{2}
		\bar{\psi} \Gamma^\mu [ A_\mu, \psi ]
	\Biggr] \,,
\end{equation}
where $A_\mu$ and $\psi$ are both $N \times N$ Hermite matrices, 
and they are a $\SO(10)$ vector and a left-handed spinor, respectively. 
We choose the scale of $A_\mu$ and $\psi$ so that the action 
takes the above form. 
$\lambda$ is a coupling constant\footnote{%
The coupling constant $\lambda$ is related to the Yang-Mills coupling 
$g_0$ by $\sqrt\lambda = \sqrt{g_0^{\protect\phantom{0}2}N}$, 
when the IIB matrix model 
is seen as the dimensional reduction of 10-dimensional supersymmetric 
$\SU(N)$ Yang-Mills theory to zero volume limit. 
We are considering the large-$N$ limit with $\lambda$ fixed to $O(1)$.
}. 
Though $\lambda$ can be absorbed by the rescaling of the fields, 
it is instead kept at first and it will later be set to 1.

The IMFA prescription is applied to the IIB matrix model by the 
following steps \cite{Nishimura:2001sx,Kawai:2002jk}. 
We first introduce the quadratic term as:
\begin{equation}
	S_0 = N\,{\rm Tr}\,\Biggl[
		\frac{1}{2} M_{\mu\nu}\,A_\mu A_\nu
		+ \frac{1}{2} m_{\mu\nu\rho}\,
		\bar{\psi} \Gamma^{\mu\nu\rho} \psi
	\Biggr] \,,
\end{equation}
which is chosen to be of most generic $\SU(N)$ invariant form. 
$M_{\mu\nu}$ and $m_{\mu\nu\rho}$ are arbitrary parameters. 
The former is symmetric with the exchange of $\mu$ and $\nu$, 
while the latter is totally anti-symmetric with $\mu$, $\nu$, and $\rho$. 

The original action $S$ is transformed nominally by adding and 
subtracting $S_0$ as 
\begin{equation}
	S \longrightarrow S_0 + (S - S_0) \,. 
\end{equation}
Then the term $(S - S_0)$ would be viewed as an interaction term 
and the perturbative expansion is constructed 
by considering the term $S_0$ as a free part. 
Instead, we consider the deformed action $S^\prime = S_0 + S$ and 
formulate the perturbation theory with reference to the coupling 
constant $\lambda$. 
Next we shift the parameters as follows by introducing the 
formal expansion parameter $g$: 
\begin{align}
	\lambda & \longrightarrow g \lambda \,,  \nonumber \\
	M_{\mu\nu} & \longrightarrow M_{\mu\nu} - g M_{\mu\nu} \,,
 \nonumber \\
	m_{\mu\nu\rho} & \longrightarrow m_{\mu\nu\rho} - g
 m_{\mu\nu\rho} \,, 
\label{eq:imp}
\end{align}
If $g$ is na\"{\i}vely taken to be 1, the fictitious parameters vanish 
and the action $S^\prime$ returns to the original action $S$. 
We reorganize the series in terms of $g$, disregard $O(g^{n+1})$ 
terms, and then set $g$ to 1. 
Thus we obtain the {\em improved} series of order $n$. 

The artificial parameters, $M_{\mu\nu}$ and $m_{\mu\nu\rho}$, are 
determined according to the principle that the result should be 
least sensitive to those parameters, since the original model 
does not rely on them. 
The dependence is brought in due to the truncation at finite orders. 
Thus, we assume if there exists a region of parameters in which 
the physical quantity such as free energy becomes stable, 
the dependence on the artificial parameters should vanish effectively, 
and the {\em true} value would be reproduced. 
This consistency condition is called plateau condition. 
The emergence of plateau is a key feature to recognize whether or not 
the IMFA prescription works well. 

Typically, the improved series of a finite order forms a flat region 
in which it fluctuates gently and accompanies a number of extrema. 
So, we adopt a criterion for identifying the plateau by the 
accumulation of extrema of the improved free energy. 
Furthermore, we estimate the values of the physical quantities 
at those extrema as the representatives of the estimates on the plateau. 
If the improved series were convergent, they would provide 
a good approximate value at high enough orders of the IMFA analysis. 
The values obtained along the IMFA prescription are considered 
to be non-perturbative, 
although the original series are based on perturbative expansions 
about a perturbative vacuum \cite{Aoyama:2005nd}. 
We can obtain a solution that corresponds to 
the non-perturbative vacuum of spontaneously broken symmetry.

\section{Ansatz}

In the case of the IIB matrix model, the total number of artificial 
parameters are quite large, namely, 
10 real numbers for $M_{\mu\nu}$ (assumed to be diagonalized by 
$\SO(10)$ rotation), and 
120 for $m_{\mu\nu\rho}$. 
It will demand an enormous effort to search for the plateau 
in this vast space of parameters. 
Therefore we impose restrictions on the configuration by 
considering symmetry that remains unbroken to diminish the number of 
parameters. 

In the former works 
\cite{Nishimura:2001sx,Kawai:2002jk,Kawai:2002ub,eighth}, 
the configurations called $\SO(d)$ ansatz 
have been intensively examined which preserve $\SO(d)$ 
rotational symmetry. 
In a practical sense, we impose the condition for the two-point 
functions of bosonic and fermionic fields:
\begin{align}
	\langle (A_{\mu})_{ij} (A_{\nu})_{kl} \rangle 
	&= 
	\frac{1}{N} C_{\mu\nu} \delta_{il} \delta_{jk} \,, \\
	\langle (\psi_{\alpha})_{ij} (\psi_{\beta})_{kl} \rangle 
	&=
	\frac{1}{\sqrt{120}N} u_{\mu\nu\rho} ({\cal C}\Gamma^{\mu\nu\rho})_{\alpha \beta} 
	\delta_{il}\delta_{jk} \,.
\end{align}
The guideline of choice is described as follows. 
First, $\SO(d)$ subgroup of $\SO(10)$ is fixed to 
which directions the expectation values of fermionic two-point 
function $u$ are zero. 
$d$ is chosen from 1 to 9. 
Toward the rest of the directions, $u$ may have non-zero value. 
Since $u$ is a rank three anti-symmetric tensor, a single non-zero 
component of $u_{\mu\nu\rho}$ is accompanied by three-dimensional 
subspace if the rotational symmetry is taken into account. 
Thus $(10-d)$ dimensional part would naturally be decomposed 
into multiples of three-dimensional blocks. 
Furthermore, those blocks are subjected to the permutation 
symmetry of the interchange of each other. 

Among those choice shown above, $d=4$ and $d=7$ cases are 
relevant; it is reported 
\cite{Kawai:2002jk}
that $d=5, 6$ cases reduce to 
$\SO(7)$ ansatz, while $d=2, 3$ cases to $\SO(4)$ 
ansatz and $d=1$ case has no solution. 

The preserved symmetry and the explicit forms of the exact 
propagators for $d=4$ and $d=7$ cases are given as follows.

\noindent
{\bf SO(7) ansatz: } 
$\SO(7) \times \SO(3)$
\begin{equation}
	C_{\mu\nu} = {\rm diag}\bigl(\ 
		\text{7$c_1$'s},\ \text{3$c_2$'s}\ 
	\bigr) \,,
	\quad
	\sla{u} = u\,\Gamma^{8,9,10} \,,
\end{equation}

\noindent
{\bf SO(4) ansatz: } 
$\SO(4) \times \SO(3) \times \SO(3) \times Z_2$
\begin{equation}
	C_{\mu\nu} = {\rm diag}\bigl(\ 
		\text{4$c_1$'s},\ \text{6$c_2$'s}\ 
	\bigr) \,,
	\quad
	\sla{u} = \frac{u}{\sqrt{2}} 
		\bigl( \Gamma^{5,6,7} + \Gamma^{8,9,10} \bigr) \,,
\end{equation}
The $Z_2$ factor stands for the permutation symmetry between 
two $\SO(3)$ factors. 

Now we extend the ansatz to incorporate larger class of 
parameter space by relaxing some restrictions above. 
In this report we consider the configuration called ``4-3-3'' ansatz, 
in which the symmetry, 
$\SO(4) \times \SO(3) \times \SO(3)$, 
should be preserved. 
It is obtained by disregarding the permutation symmetry of 
$\SO(4)$ ansatz. 
The forms of the exact propagators are taken as follows: 
\begin{equation}
C_{\mu\nu} = \left(
	\begin{array}{ccc|ccc|ccc}
	c_1 & & & & & \\
	 & \ddots %
	\begin{picture}(0,0)\put(-2,8){\rotatebox{52}{\raisebox{0pt}[0pt][0pt]{\makebox[0pt]{$\left.\rule{0pt}{30pt}\right\}$}}}}\put(0,12){4}\end{picture}%
& & & & \\
	 & & c_1 & & & \\
	\hline 
	 & & & c_2 & & \\
	 & & & & \ddots %
	\begin{picture}(0,0)\put(-2,8){\rotatebox{52}{\raisebox{0pt}[0pt][0pt]{\makebox[0pt]{$\left.\rule{0pt}{30pt}\right\}$}}}}\put(0,12){3}\end{picture}%
 & \\
	 & & & & & c_2 \\
	\hline
	 & & & & & & c_3 \\
	 & & & & & & & \ddots %
	\begin{picture}(0,0)\put(-2,8){\rotatebox{52}{\raisebox{0pt}[0pt][0pt]{\makebox[0pt]{$\left.\rule{0pt}{30pt}\right\}$}}}}\put(0,12){3}\end{picture}%
 \\
	 & & & & & & & & c_3 \\
	\end{array}
\right) \,,
\quad
\sla{u} = 
	u_1\,\Gamma^{5,6,7} + u_2\,\Gamma^{8,9,10} \,,
\end{equation}
The three-dimensional blocked form is still respected in the present 
choice so that the number of non-zero fermionic flux should be 
kept small. 

This configuration reduces to $\SO(4)$ ansatz or $\SO(7)$ 
ansatz in their respective limits. 
\begin{equation}
\begin{aligned}
	c_2 = c_3 \,,\quad u_1 = u_2\ \left(= \frac{u}{\sqrt{2}}\right)
	\quad & \Leftrightarrow \quad 
	\text{$\SO(4)$ ansatz,} \\
	c_1 = c_2 \,,\quad u_1 = 0
	\quad & \Leftrightarrow \quad 
	\text{$\SO(7)$ ansatz,} \\
\end{aligned}
\end{equation}
It has been argued by comparing the free energy, which of 
$\SO(d)$ ansatz dominates in the configuration space, 
where the values of free energy are estimated individually 
for each ansatz. 
Since the 4-3-3 ansatz covers both $\SO(4)$ and $\SO(7)$ 
ansatz as subsets, the comparison will now be made on the 
same basis.

\section{Results}

We applied the IMFA method to the IIB matrix model and obtained the 
improved series of free energy for 4-3-3 ansatz up to third order. 

In order to evaluate the free energy we use the fact that the free
energy $F$ is 
related by Legendre transformation to 2PI free energy $G$ which 
is given by the sum of two-particle irreducible diagrams, in 
terms of exact propagators, $C_{\mu\nu}$ and $\sla{u}$ \cite{Fukuda:1995im}. 
This is mainly technical reason, for the number of diagrams 
reduces drastically by working with 2PI diagrams. 

Since the large-$N$ limit of the model is considered, the 
dominant contribution derives from the planar diagrams
\cite{Brezin:1977sv}, 
which we have only to evaluate. 
The number of 2PI planar vacuum diagrams are, 2 at zeroth, first 
and second order, and 4 at third order. 
The explicit expression of free energy up to third order is 
presented in the appendix. 

From the 2PI free energy $G(c,u)$, the free energy $F(M,m)$ 
is obtained by: 
\begin{equation}
	F(M,m) = \left.\left\{ 
	G(c,u) 
	+ \frac{4}{2} M_1 c_1 + \frac{3}{2} M_2 c_2 + \frac{3}{2} M_3 c_3 
	- \frac{8}{2}\sum_{i=1,2} m_i u_i 
	\right\}\right|_{c=c(M,m),\,u=u(M,m)} \,,
\end{equation}
where $c_i (i=1,2,3)$ and $u_j (j=1,2)$ are determined by the solution 
of the following relations: 
\begin{equation}
	M_i = \frac{\partial G(c,u)}{\partial c_i} \,,
	\quad
	m_j = \frac{\partial G(c,u)}{\partial u_j} \,. 
\end{equation}

Next we performed the IMFA prescriptions to the free energy and 
obtained the improved series as a function of $\{M_i\}$ and $\{m_j\}$ by
the transformation (\ref{eq:imp}) and setting $g$ equal to 1. 
In order to determine those artificial parameters we search for 
the solution of plateau condition by identifying the accumulation 
of the extrema of improved series. 
The extrema are found by numerical means. 

\begin{figure}
\includegraphics[scale=1.0]{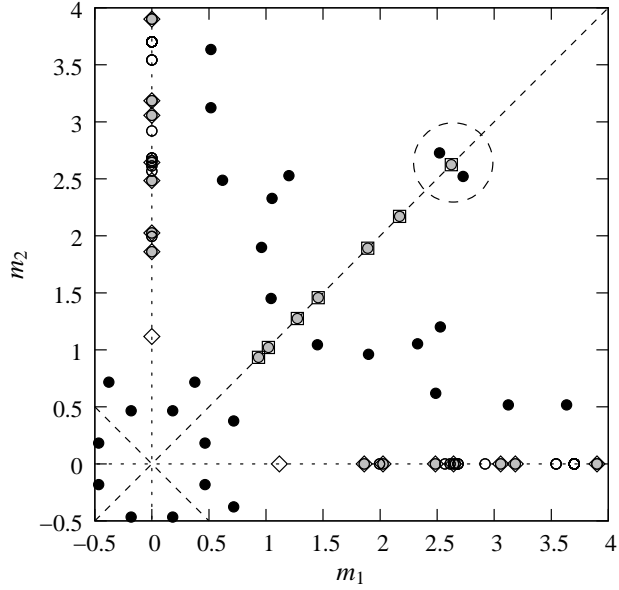}
\caption{Distribution of extrema of the improved free energy for 4-3-3 ansatz 
plotted on $m_1$-$m_2$ plane (bullets). 
Vertical and horizontal lines correspond to $\SO(7)$ ansatz, and 
a diagonal line does to $\SO(4)$ ansatz.}
\label{fig:extrema}
\end{figure}

Fig.~\ref{fig:extrema} shows the distribution of extrema of 
the improved free energy for 4-3-3 ansatz plotted on $m_1$-$m_2$ plane. 
The extrema are shown by bullets 
(filled $\bullet$, shaded \textcolor{gray}{$\bullet$}, and unfilled $\circ$). 
Vertical and horizontal lines ($m_1=0$ or $m_2=0$) 
correspond to $\SO(7)$ ansatz, and 
a diagonal line ($m_1=m_2$) does to $\SO(4)$ ansatz. 
The shaded bullets (\textcolor{gray}{$\bullet$}) on the diagonal line 
represent the $\SO(4)$-preserving solutions ($M_2=M_3$). 
The shaded bullets (\textcolor{gray}{$\bullet$}) on the vertical 
and horizontal lines represent the $\SO(7)$-preserving solutions 
which satisfy $M_1 = M_2$ (or $M_1 = M_3$), 
while the circles ($\circ$) correspond to the solutions in which 
$\SO(7)$ symmetry is not preserved, i.e. $M_1 \ne M_2$ (or $M_1 \ne M_3$). 

There are also plotted the extrema of $\SO(4)$ ansatz (squares $\Box$) 
and those of $\SO(7)$ ansatz (diamonds $\Diamond$) for reference. 
It is seen that some of the $\SO(7)$ solutions ($\Diamond$) coincide 
with those of 4-3-3 ansatz (\textcolor{gray}{$\bullet$}) 
on vertical or horizontal lines. 
The other solutions disappear from the solutions of 4-3-3 ansatz, 
which are considered to be unstable in the sense that 
they correspond to saddle points in extended space of parameters. 
It is consistent that $\SO(7)$-nonpreserving solutions ($\circ$) 
do not coincide with the solutions of $\SO(7)$ ansatz. 
Similar consideration applies to the extrema on the diagonal line. 
The $\SO(4)$ solutions ($\Box$) that appear as extrema of 
4-3-3 ansatz (\textcolor{gray}{$\bullet$}) are considered to be 
stable, while the others are to be unstable. 
This speculation may lead to the prospect that the extended 
parameter space works to distinguish the plausible plateau 
from the others. 

It is yet unclear what configuration would become dominant at 
such low order of calculations. 
However, there seems to exist a non-trivial accumulation of 
extrema on and near the $\SO(4)$ symmetric subspace 
(enclosed by dashed circle in the figure). 
The $\SO(4)$ symmetric extrema in this region corresponds 
to that of $\SO(4)$ ansatz case which is known to belong 
to the would-be plateau in higher order analysis. 
It is found that 
each of those two extrema located near the diagonal line 
has the property $M_2 \simeq M_3$ for the bosonic variables as well.
It implies that $\SO(4)$ symmetry is almost restored in this region. 
This supports that the region would be hopeful for plateau. 
To clarify the situation, higher order contributions for 
the 4-3-3 ansatz will be required as a future outlook. 

\begin{acknowledgments}
Y.~S. is supported by the Special Postdoctoral Researchers Program 
at RIKEN. 
\end{acknowledgments}

\appendix
\section{2PI free energy of 4-3-3 ansatz}

Here we present the two-particle irreducible free energy for 
4-3-3 ansatz up to third order. 
The additive constant is adjusted to the definition in 
\cite{Nishimura:2001sx}.

\begin{align}
G / N^{2} & = 
3 ( 1 + \log 2 ) 
- \frac{1}{2} \log c_1^{\ 4} c_2^{\ 3} c_3^{\ 3} 
+ 8 \frac{1}{2} \log( u_1^{\ 2} + u_2^{\ 2} ) 
\nonumber \\ 
& + \lambda \Biggl\{
6 c_1^{\ 2} 
+ 3 c_2^{\ 2} 
+ 3 c_3^{\ 2} 
+ 24 c_2 u_1^{\ 2} 
+ 12 c_1 c_2 + 12 c_1 c_3 + 9 c_2 c_3 
\nonumber \\ & \quad\quad 
- 24 c_3 u_1^{\ 2} 
- 32 c_1 u_1^{\ 2} 
- 24 c_2 u_2^{\ 2} 
- 32 c_1 u_2^{\ 2} 
+ 24 c_3 u_2^{\ 2} 
\Biggr\} 
\nonumber \\
& + \lambda^{2} \Biggl\{
- 32 c_1^{\ 2} u_1^{\ 4} 
- \frac{27}{2} c_2^{\ 2} c_3^{\ 2} 
- 12 c_2^{\ 2} u_1^{\ 4} 
- 12 c_3^{\ 2} u_1^{\ 4} 
- 18 c_1^{\ 2} c_3^{\ 2} 
- 18 c_1^{\ 2} c_2^{\ 2} 
\nonumber \\ & \quad\quad 
- 96 c_1 c_2 u_1^{\ 4} 
- 96 c_1 c_3 u_1^{\ 4} 
- 72 c_2 c_3 u_1^{\ 4} 
- 9 c_1^{\ 4} 
- \frac{9}{2} c_2^{\ 4} 
- \frac{9}{2} c_3^{\ 4} 
- 32 c_1^{\ 2} u_2^{\ 4} 
\nonumber \\ & \quad\quad 
- 64 c_1^{\ 2} u_1^{\ 2} u_2^{\ 2} 
- 96 c_1 c_2 u_2^{\ 4} 
- 192 c_1 c_2 u_1^{\ 2} u_2^{\ 2} 
- 96 c_1 c_3 u_2^{\ 4} 
- 192 c_1 c_3 u_1^{\ 2} u_2^{\ 2} 
\nonumber \\ & \quad\quad 
- 12 c_3^{\ 2} u_2^{\ 4} 
+ 72 c_3^{\ 2} u_1^{\ 2} u_2^{\ 2} 
+ 432 c_2 c_3 u_1^{\ 2} u_2^{\ 2} 
- 72 c_2 c_3 u_2^{\ 4} 
- 12 c_2^{\ 2} u_2^{\ 4} 
+ 72 c_2^{\ 2} u_1^{\ 2} u_2^{\ 2} 
\Biggr\} 
\nonumber \\ 
& + \lambda^{3} \Biggl\{
- 96 c_2^{\ 4} u_1^{\ 4} 
- 576 c_1 c_2 c_3 u_2^{\ 6}
- 1728 c_1 c_2 c_3 u_1^{\ 2} u_2^{\ 4}
- 1728 c_1 c_2 c_3 u_1^{\ 4} u_2^{\ 2}
- 96 c_1 c_3^{\ 2} u_2^{\ 6} 
\nonumber \\ & \quad\quad 
- 672 c_1 c_3^{\ 2} u_1^{\ 2} u_2^{\ 4}
- 1056 c_1 c_3^{\ 2} u_1^{\ 4} u_2^{\ 2}
- 96 c_2^{\ 4} u_2^{\ 4}
+ 576 c_2^{\ 4} u_1^{\ 2} u_2^{\ 2}
- 24 c_2^{\ 3} u_2^{\ 6} 
\nonumber \\ & \quad\quad 
- 1944 c_2^{\ 3} u_1^{\ 2} u_2^{\ 4}
- 96 c_3^{\ 4} u_1^{\ 4}
+ 72 c_2^{\ 2} c_3 u_2^{\ 6}
- 1080 c_2^{\ 2} c_3 u_1^{\ 2} u_2^{\ 4}
+ 1080 c_2^{\ 2} c_3 u_1^{\ 4} u_2^{\ 2} 
\nonumber \\ & \quad\quad 
- 72 c_2 c_3^{\ 2} u_2^{\ 6}
+ 1080 c_2 c_3^{\ 2} u_1^{\ 2} u_2^{\ 4}
- 1080 c_2 c_3^{\ 2} u_1^{\ 4} u_2^{\ 2}
- 96 c_3^{\ 4} u_2^{\ 4}
+ 576 c_3^{\ 4} u_1^{\ 2} u_2^{\ 2} 
\nonumber \\ & \quad\quad 
- 232 c_3^{\ 3} u_2^{\ 6}
+ 1176 c_3^{\ 3} u_1^{\ 2} u_2^{\ 4}
- 1944 c_3^{\ 3} u_1^{\ 4} u_2^{\ 2}
+ 14 c_2^{\ 6}
- 128 c_1^{\ 3} u_1^{\ 6}
+ 39 c_2^{\ 3} c_3^{\ 3} 
\nonumber \\ & \quad\quad 
- 232 c_2^{\ 3} u_1^{\ 6} 
- 24 c_3^{\ 3} u_1^{\ 6}
+ 36 c_1^{\ 2} c_2^{\ 2} c_3^{\ 2}
+ 52 c_1^{\ 3} c_2^{\ 3}
+ 52 c_1^{\ 3} c_3^{\ 3}
- 384 c_1^{\ 2} c_3^{\ 2} u_1^{\ 4} 
\nonumber \\ & \quad\quad 
- 768 c_1^{\ 2} c_3 u_1^{\ 6} 
- 576 c_1 c_2 c_3 u_1^{\ 6}
- 96 c_1 c_2^{\ 2} u_1^{\ 6}
- 480 c_1 c_3^{\ 2} u_1^{\ 6}
- 72 c_2^{\ 2} c_3 u_1^{\ 6} 
\nonumber \\ & \quad\quad 
+ 72 c_2 c_3^{\ 2} u_1^{\ 6}
+ 30 c_1^{\ 6}
+ 14 c_3^{\ 6} 
+ 18 c_1^{\ 4} c_3^{\ 2}
+ 18 c_1^{\ 4} c_2^{\ 2}
- 128 c_1^{\ 3} u_2^{\ 6}
- 384 c_1^{\ 3} u_1^{\ 2} u_2^{\ 4} 
\nonumber \\ & \quad\quad 
- 384 c_1^{\ 3} u_1^{\ 4} u_2^{\ 2}
+ 12 c_1^{\ 2} c_3^{\ 4} 
- 1536 c_1^{\ 2} c_2 u_1^{\ 2} u_2^{\ 4}
- 768 c_1^{\ 2} c_2 u_1^{\ 4} u_2^{\ 2}
- 384 c_1^{\ 2} c_2^{\ 2} u_2^{\ 4} 
\nonumber \\ & \quad\quad 
- 384 c_1^{\ 2} c_2^{\ 2} u_1^{\ 2} u_2^{\ 2}
- 768 c_1^{\ 2} c_2 u_2^{\ 6} 
- 192 c_1^{\ 4} u_1^{\ 4}
+ 12 c_1^{\ 2} c_2^{\ 4}
+ 9 c_2^{\ 2} c_3^{\ 4}
+ 9 c_2^{\ 4} c_3^{\ 2} 
\nonumber \\ & \quad\quad 
- 192 c_1^{\ 4} u_2^{\ 4}
- 384 c_1^{\ 4} u_1^{\ 2} u_2^{\ 2} 
- 384 c_1^{\ 2} c_3^{\ 2} u_1^{\ 2} u_2^{\ 2}
- 768 c_1^{\ 2} c_3 u_1^{\ 2} u_2^{\ 4}
- 1536 c_1^{\ 2} c_3 u_1^{\ 4} u_2^{\ 2} 
\nonumber \\ & \quad\quad 
+ 1176 c_2^{\ 3} u_1^{\ 4} u_2^{\ 2}
- 480 c_1 c_2^{\ 2} u_2^{\ 6} 
- 1056 c_1 c_2^{\ 2} u_1^{\ 2} u_2^{\ 4}
- 672 c_1 c_2^{\ 2} u_1^{\ 4} u_2^{\ 2}
\Biggr\} 
\nonumber \\
& + O( \lambda^{4} ) 
\end{align}


\end{document}